\documentclass{aastex631}

\usepackage{natbib}
\usepackage{xcolor}
\newcommand{\eps}[1]{\mbox{log~$\epsilon$(#1)}} 
\newcommand\species[2]{#1 {\sc #2}}
\newcommand\iso[2]{$^{\rm #1}$#2}

\def\eg{\mbox{e.g.}}

\def\teff{\mbox{$T_{\rm eff}$}}
\def\logg{\mbox{log~{\it g}}}
\def\kmsec{\mbox{km~s$^{\rm -1}$}}

\def\vsini{$V$sin($i$)}

\def\vmicro{$V_{mic}$}
\def\carbiso{{$^{12}$C/$^{13}$C}}

\shorttitle{HPF Chemical Compositions}
\shortauthors{Nagarajan et al.}

\begin{document}

\title{CHEMICAL COMPOSITIONS OF RED GIANT STARS IN THE OLD OPEN
       CLUSTER NGC 7789}

\correspondingauthor{Neel Nagarajan}
\email{neelnagarajan@utexas.edu}

\author[0000-0002-7112-2086]{Neel Nagarajan}
\affiliation{Department of Astronomy and McDonald Observatory,
             The University of Texas, Austin, TX 78712, USA;
             neelnagarajan@utexas.edu}
\author[0000-0002-3456-5929]{Christopher Sneden}
\affiliation{Department of Astronomy and McDonald Observatory,
             The University of Texas, Austin, TX 78712, USA;
             chris@verdi.as.utexas.edu}
\author[0000-0002-2516-1949]{Melike Af\c{s}ar}
\affiliation{Department of Astronomy and Space Sciences,
             Ege University, 35100 Bornova, \.{I}zmir, Turkey;
             melike.afsar@gmail.com}
\author[0000-0002-3007-206X]{Catherine A. Pilachowski}
\affiliation{Department of Astronomy, Indiana University,
             Indiana University, Bloomington, IN 47405, USA
             cpilacho@indiana.edu}

\begin{abstract}

We have gathered optical-region spectra, derived model atmosphere parameters, 
and computed elemental abundances for 15 red giant stars in the open cluster 
NGC 7789. 
We focus on the light element group CNOLi that provides clues to 
evolutionary changes associated with internal fusion events and chemical mixing.
We confirm and extend an early report that NGC 7789 stars 193 and 301 have 
anomalously large Li abundances, and that these values are apparently 
unconnected to any other elements' abundances in these stars. 
A companion study of \species{He}{i} $\lambda$10830 lines in both field 
stars and cluster members shows that star 301 has a strong He feature 
while star 193 does not. 
Possible explanations for the large Li abundances of these stars include
helium flash-induced mixing events and binary interactions at some past
or present times.
In either case an internal eruption of energy could cause fresh synthesis of 
lithium via the Cameron-Fowler Berillyum transport mechanism. 
Rapid transport of lithium to the outer layers may have created significant 
chromospheric transient disturbances, producing enough helium ionization to 
allow for the strong $\lambda$10830 absorption in star 301.

\end{abstract}

\section{INTRODUCTION}\label{intro}

Red giant stars typically exhibit low lithium abundances relative to those of 
main sequence stars. 
During main sequence stars' lives, Li is easily destroyed in high temperature 
interior regions as part of the proton-proton cycle of hydrogen fusion:  
\iso{7}{Li}($p,\alpha$)$\rightarrow$\iso{4}{He}. As main sequence stars evolve to become subgiants and then red giants, their deepening convective envelopes mix surface and interior layers, effectively cleaning the stellar atmospheres of their natal Li contents.

Because of this lithium-depletion process we typically observe surface
Li abundances to be large in main sequence stars 
(as much as \eps{Li}~$\sim$~3.3)\footnote{
We adopt the standard spectroscopic notation
\citep{wallerstein59} that for elements A and B,
[A/B] $\equiv$ log$_{\rm 10}$(N$_{\rm A}$/N$_{\rm B}$)$_{\star}$ $-$
log$_{\rm 10}$(N$_{\rm A}$/N$_{\rm B}$)$_{\odot}$.
We use the definition
\eps{A} $\equiv$ log$_{\rm 10}$(N$_{\rm A}$/N$_{\rm H}$) + 12.0, and
equate metallicity with the stellar [Fe/H] value.}
but much lower in red giant stars (\eps{Li}~$\lesssim$~1.5).
However, around one percent of red giant stars in our Galaxy have unusually 
high Li abundances, as evidenced by a strong Li $\lambda$6707 absorption. 
Many papers have reported discoveries of Li-rich giants. 
Lithium excess is defined in various ways, but most papers suggest that
red giants have anomously large Li abundances if \eps{Li}~$\gtrsim$~1.5.

Two main sources have been suggested for such Li overabundances in 
red giants.  
First, it might be possible for Li to be added as an expanding stellar
envelope engulfs a companion, typically thought to be a terrestrial
companion (e.g., \citealt{alexander67}).  
Second, fresh Li could be produced
in the interiors of a star in the
so-called beryllium transport mechanisim \citep{cameron71}, in which Li is
synthesized via $^3$He($\alpha,\gamma$)$^7$Be(e$^-$,$\nu$)$^7$Li and then
dredge up to the surface before it can be destroyed in normal pp-chain hydrogen
fusion.
The assets and liabilities of these ideas are discussed in more detail in
many papers, e.g., \cite{casey19}, \cite{deepak19}, \cite{martell21}.
Recently, \cite{sneden22} (hereafter Paper 1)
conducted a large survey of \species{He}{i} 10830~\AA\ chromospheric lines,
discovering a correlation between high Li abundances and very large 
$\lambda$10830 absorption strengths.
Apparently the generation of Li in red giant envelopes is often accompanied by
major disturbances in their outer atmospheres.

Some recent studies have begun to search for 
Li-rich stars in open clusters, where masses and ages can be estimated; e.g., 
\cite{anthonytwarog13}, \cite{carlberg16}, \cite{magrini21}, and \cite{sun22}.
In the present work we revisit the intermediate age open star cluster NGC~7789,
which contains the first two red giants reported to have anomalously large Li abundances (\eps{Li} $\sim$ +2.4, \citealt{pilachowski86}). 
This cluster is relatively bright and has an extensive literature history.
At age 1.5~Gyr \citep{gao18}\
NGC~7789 has a well-developed red giant branch, and has been subjected to
several high-resolution spectroscopic analyses, but apparently without focus
on its Li abundance anomaly since the \citeauthor{pilachowski86} study.
Basic data for NGC~7789 are summarized in Table~\ref{tab-cluster}.

\cite{pilachowski86} reported significant Li abundance excesses in 
NGC~7789 193 and NGC~7789 301 (hereafter labeled stars 193 and 301).
Paper 1 included these stars and 8 other cluster members from the
\citeauthor{pilachowski86} sample.
However, the optical spectra from the paper over 3 decades ago
were of modest spectral resolution and signal-to-noise ratio 
(SNR); these
stars lack recent accessible optical high-resolution spectra
in the \species{Li}{i} $\lambda$6707 region. 
Therefore we have gathered and analyzed new optical echelle spectra for these NGC 7789 stars.
In this paper we report Li abundances and discuss the Li-He connection for 
NGC~7789 red giants.

In \S\ref{obsred} we describe the stellar sample, our observations, 
and the reduction steps to produce the final spectra.
The model atmosphere and abundance analyses are outlined in \S\ref{analyses}.
We discuss the results in \S\ref{discussion} with particular emphasis on the
LiCNO abundance group and relation to the $\lambda$10830 data from Paper~1,
and summarize our conclusions in \S\ref{conclusions}.

\vspace*{0.2in}
\section{OBSERVATIONS AND REDUCTIONS}\label{obsred}

\subsection{The Stellar Sample}\label{stars}

We observed and analyzed 15 NGC~7789 red giants.
The program stars are listed in Table~\ref{tab-stars} along with their
astrometric and photometric properties of interest to our work. 
Inspection of the parallax and proper motion data in this table reveals
that star 193 is discordant and likely not
a member of NGC~7789.
The possibility that star 193 is a non-member arose in \cite{pilachowski86},
but the star
was not discarded because its high Li abundance would make the
star interesting regardless of its membership status.
We also retained star 193 in our work, and will comment on it later.

We began with a cluster membership analysis.
We considered stars in a search radius roughly double the radius of the WEBDA cluster chart for NGC 7789. 
We gathered the astrometric data for stars in this
area from the GAIA \citep{GAIA16} data archive \cite{GAIA22}. 
We retained stars with GAIA parameters 
parallax $p$ and proper motions $\mu_{RA}$, $\mu_{Dec}$
in the following ranges respectively:
0.4 to 0.5, $-$1.2 to $-$0.6, and $-$2.3 to $-$1.65. 
All these ranges are centered on the cluster mean values as presented in SIMBAD.

After elimination of non-member stars, we then constructed an 
extinction-corrected Gaia-based color magnitude diagram (CMD) for NGC~7789.
We converted $E(B-V)$ from Table~\ref{tab-cluster} to $E(BP-RP)$~=~0.39 
using two equations: $E(BP-RP)$~$\simeq$~4.507$EW_{862}$ (developed by 
\citealt{schultheis22}), and $E(B-V)$~$\simeq$~3.1$EW_{862}$ (from table 3 of  
\citealt{schultheis22}). 
Here, $EW_{862}$ is the equivalent width of the diffuse interstellar band 
at 862 nm and $E(B-V)$~$\simeq$~3.1$EW_{862}$ is a relationship between 
the band strength and the $UBV$ standard reddening value.

\begin{figure}
\epsscale{0.90}
\plotone{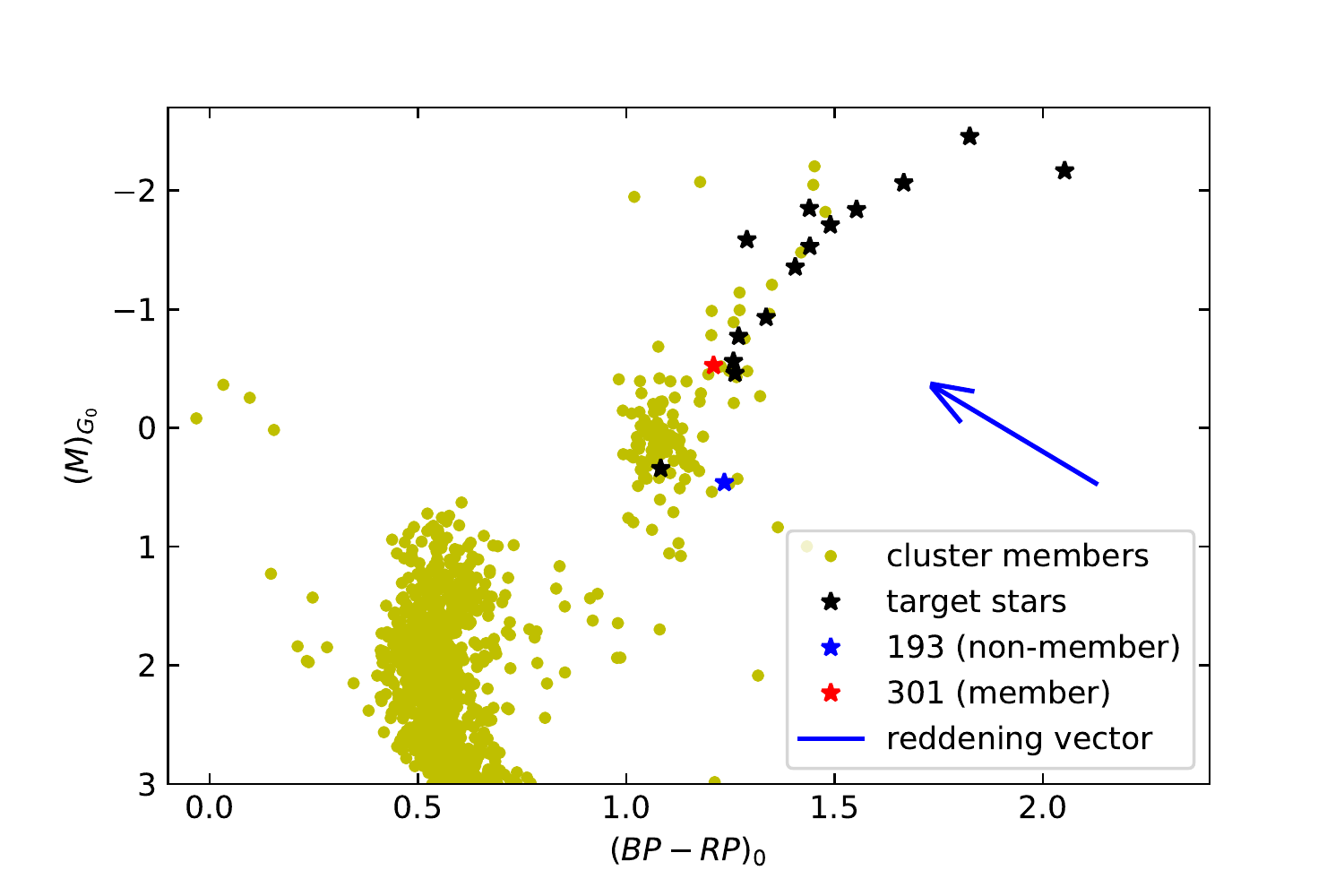}
\caption{\label{fig1}
\footnotesize
   An extinction-corrected color magnitude diagram (CMD) for NGC 7789 using 
   GAIA photometric data.
   As identified in the plot legend, the star-shaped symbols show
   the red giants of this study. 
   The reddening vector shows the direction and length by which we shifted 
   the stars on the CMD to correct for interstellar reddening effects. }
\end{figure}

We converted the total extinction to Gaia units via the relation 
$A_G/A_V$~$\simeq$~0.95 
from \cite{jordi10}.
Application of these formulae yielded the $(BP-RP)_0$ and $M_{G0}$ values
employed to create
the CMD shown in Figure~\ref{fig1}.
This is similar to other CMDs for NGC~7789 in the literature, e.g. 
\cite{overbeek15}, \cite{gao18}, and \cite{cantatgaudin18}. 
We have included star 193 in this plot even though it is not a 
probable NGC 7789 member, and we have arbitrarily
used the cluster reddening values in placing it in the figure.
Its position in Figure~\ref{fig1} should not be used to infer 
its evolutionary history. 
Furthermore, star 765's membership 
may also be in question due to it's relatively low parallax 
(Table~\ref{tab-stars}), but its proper motions are consistent with
NGC~7789 membership.
The dereddened CMD presented in Figure~5 of \cite{sandquist20} for 
multiple open clusters shows that the approximate red clump center is 
at [$(BP-RP)$,$M_{G0}$] = [1.1,0.4], 
which is essentially coincident to the red clump 
center for NGC 7789.

In Figure~\ref{fig1} the stars with $(BP-RP)_0$~$\lesssim$~0.4 
are easily identified as blue straggler stars (BSS).  
Such members of NGC~7789 have been known for many decades since the
pioneering CMD study of \cite{burbidge58}.
Blue stragglers probably result from mass transfer in binary or
triple systems (\eg, \citealt{mccrea64}, \citealt{hills76}, 
\citealt{perets09}), suggesting that NGC~7789 has a relatively large set
of present or past multiple star systems.

\subsection{High-Resolution Spectra}\label{spectra}

We gathered high resolution spectra of the target stars with the 
McDonald Observatory 2.7m Harlan J. Smith Telescope and Tull Echelle Spectrograph 
\cite{tull95}.  
The combination of slit, dispersing elements, and detector yielded spectral 
resolution of $R~\equiv \lambda/\Delta\lambda$ $\simeq$~45,000.
The observed spectral range covered 3500$-$9500~\AA. 
However, our red giants have sharply decreasing fluxes and substantial
spectral line blanketing toward shorter wavelengths, so in consequence our
analyses considered only wavelength regions with $\lambda$~$>$~5300~\AA.
Typical signal-to-noise ratios near 6000~\AA\ were SNR~$\sim$~100 per
resolution element.
The obtained spectra included star 971, a red giant tip member of NGC~7789.
This star is by far the reddest one of our sample (Table~\ref{tab-stars}).
Its
low temperature along with inclement weather during observation resulted in a very noisy
reduced spectrum, and so we dropped it from further consideration in
this paper.

\begin{figure}
\epsscale{0.50}
\center
\includegraphics[angle=-0,width=0.7\textwidth]{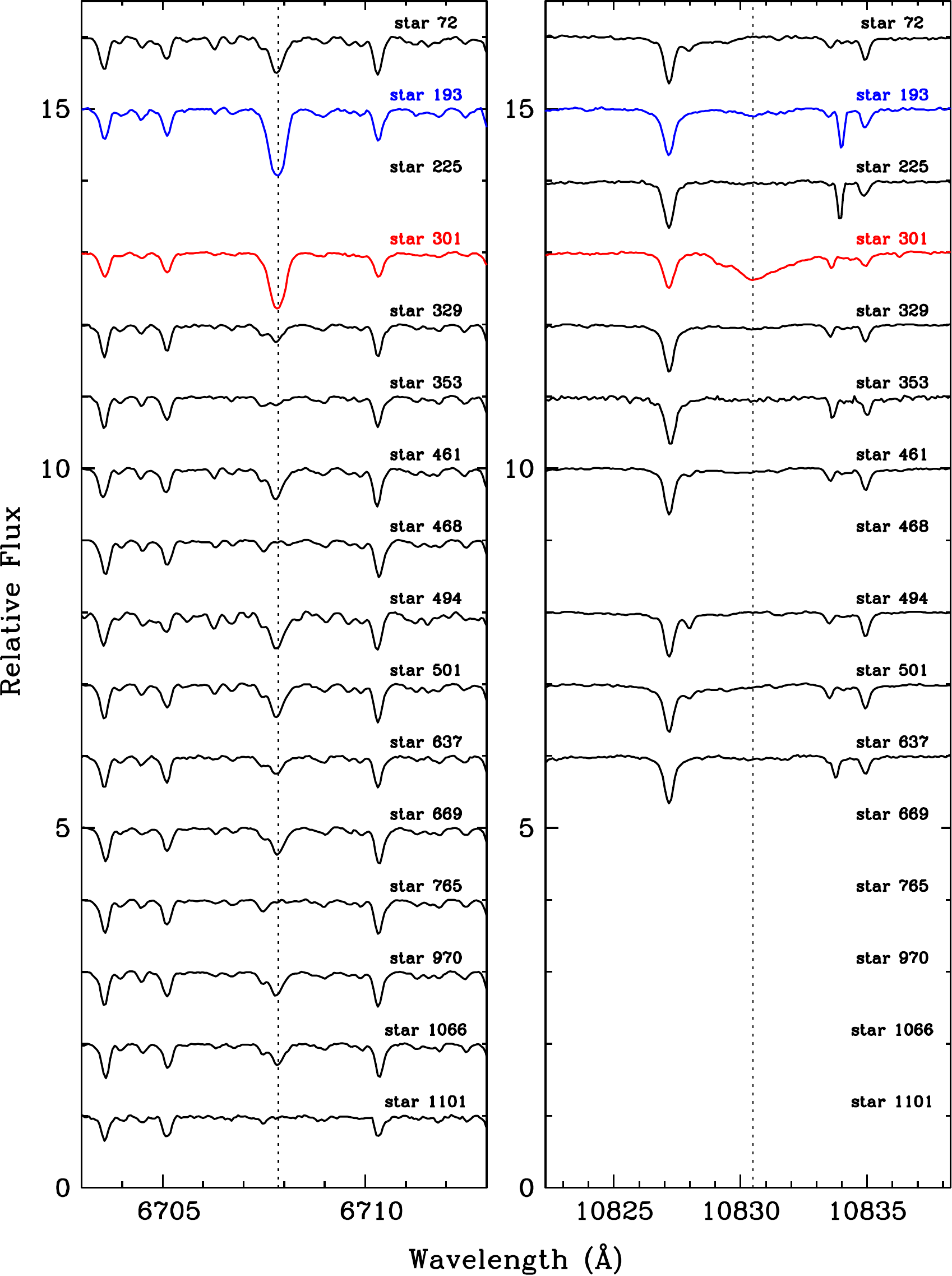}
\caption{\label{fig2}
\footnotesize
   Spectra of the \species{Li}{i} $\lambda$6707 resonance transition in all 
   NGC 7789 red giants observed with the Tull Spectrograph (left panel)
   and of the \species{He}{i} $\lambda$10830 transition in all of this
   cluster's stars observed with 
   HET/HPF by \cite{sneden22} (right panel).
   Attention is called to Li-rich stars 193 (in blue) and 301 (in red).
   All spectra of NGC~7789 giants observed with either instrument are
   shown in this figure.
}
\end{figure}

All spectra were obtained in a single observing run covering 7 nights.
Total integration times for 
each program star
were 2$-$3 hours, obtained
in individual half-hour exposures.
The program stars' observations were accompanied by bias, dark, flat-field 
incandescent, and Th-Ar lamp integrations.
We also observed several rapidly rotating hot stars to facilitate 
cancellation of telluric features in the program stars.

\subsection{Reductions and Equivalent Widths}\label{reductions}

We performed all spectral reductions using the IRAF\footnote{
http://iraf.noao.edu/} \citep{tody86,tody93} facility.
The reduction tasks included bias subtraction, cosmic ray excision, 
echelle order extraction, continuum normalization, pixel to wavelength
conversion, and velocity correction to rest wavelength scales.

One of the chief goals of our study is to investigate the relationship 
between Li abundance and \species{He}{i} $\lambda$10830 absorption strength.
In the left-hand panel of Figure~\ref{fig2} we show spectra of 
the \species{Li}{i} $\lambda$6707 resonance feature in all stars observed here,
and in the right-hand panel we show the \species{He}{i} line in spectra
collected for Paper~1.

We measured EWs of unblended atomic lines, beginning with the line
lists used by \cite{bocek15} for their study of open cluster NGC~752 and 
by \cite{bocek16} for NGC 6940.
We calculated
the EWs with the SPECTRE\footnote{
https://www.as.utexas.edu/~chris/spectre.html}
spectrum analysis code \citep{fitzpatrick87}.
The line profiles were usually modeled by Gaussian functions, with occasional
Voigt functions for the strongest lines.
EWs could not be reliably measured for some lines below 5300~\AA\ due to 
sharply decreasing fluxes and substantial spectral line blanketing;
those lines were excluded from our analyses.
The EWs are listed in Table~\ref{tab-ews}.

\vspace*{0.2in}
\section{ABUNDANCE ANALYSES}\label{analyses}

In this section we 
describe the
steps in analyses that led to
final abundances of elements in our NGC~7789 program stars. 
Our focus is
on model atmospheric parameters and abundances of the
LiCNO element group.

\subsection{Model Atmospheres}\label{models}

We determined model atmospheric parameters and abundances from the 
line data information in the NGC~7789 spectra.
Initially we attempted to use the measured EWs to estimate model parameters
in a standard manner:
(1) \teff, from requiring low and high excitation \species{Fe}{i} lines to
yield the same Fe abundance on average; (2) \logg, from requiring 
abundance agreement between
neutral and ionized lines of Fe and Ti; (3) \vmicro, 
by forcing weak and strong \species{Fe}{i} lines 
to yield similar abundances;
and (4) [M/H] by asking for stellar model metallicity to be consistent with
derived abundances of most elements.
We tried to follow the procedures outlined in \cite{bocek15,bocek16} for 
NGC~752 and NGC~6940.

However, NGC~7789 red giant stars are substantially redder/cooler than 
those in the other two clusters, and its lines have significantly larger EWs.
Unfortunately, almost all \species{Fe}{i} lines employed for the other
clusters lie
on the flat part of the curve-of-growth
in NGC 7789 red giant stars.
In general, the strongest \species{Fe}{i} lines have low excitation energies.
The relative abundance insensitivity of these lines, combined with the
line strength and excitation correlation makes it very difficult to
determine \vmicro\ and \teff\ in the traditional manner.

Therefore we chose to derive \teff\ values purely from atomic line depth
ratios (LDRs).
This method was pioneered by \cite{gray91} and involves identification of
pairs of absorption lines whose strengths have very different responses to
variations in \teff.  
\citeauthor{gray91} considered G-K stars, and concentrated on the 6200~\AA\
spectral domain that has many low-excitation \species{V}{i} transitions that
become much stronger with decreasing \teff, usually pairing them with higher
excitation \species{Fe}{i} transitions that are much less sensitive to
\teff\ changes.
The physical principles involved are basic Boltzmann/Saha relations that
describe the populations of ionization/excitation states in stellar 
atmospheres (see extended discussions of these issues 
in \citealt{gray08}).
The work on LDRs has expanded to include other species in other spectral
regions for stars in different temperature-gravity domains, \eg,
\cite{strassmeier00}, \cite{lopezvaldivia19}.
Here we follow \cite{bocek15,bocek16} in using the LDR calibrations of
\cite{biazzo07a} which are based on the original \citeauthor{gray91} work.

With \teff\ established the other parameters \logg, \vmicro, and [M/H]
were derived from line EWs as outlined above.
In particular the microturbulent velocity was determined from forcing 
weak and strong \species{Fe}{i} lines to yield the same average abundances.
The gravity was set exclusively through the requirement to force abundance
agreement between these two Fe ionization states.
We employed interpolated models from the ATLAS\footnote{
Available at http://kurucz.harvard.edu/grids.html; model interpolation
software was kindly provided by A. McWilliam and I. Ivans.}
grid \citep{kurucz11,kurucz18}.
To compute abundances we used the 
current version of the LTE plane-parallel line analysis code
MOOG \citep{sneden73}.\footnote{
Available at http://www.as.utexas.edu/$\sim$chris/moog.html}
The derived atmospheric parameters are listed 
in Table~\ref{tab-atmospheres},
along with the abundances of \species{Fe}{i} and \species{Fe}{ii}.

The Fe abundances were determined from on average 49 \species{Fe}{i} 
and 10 \species{Fe}{ii} species.
From these
slightly sub-solar values we set a uniform metallicity of [M/H]~= $-$0.1.
The mean metallicity for NGC~7789, $<$[Fe/H]$>$~= $-$0.02 with 
$\sigma$~= 0.05, is in reasonable accord with those determined in previous 
high-resolution spectroscopic studies:
[Fe/H]~= $-$0.04, $\sigma$~= 0.05 \cite{tautvaisiene05};
[Fe/H]~= $+$0.04, $\sigma$~= 0.07 \cite{pancino10};
[Fe/H]~= $+$0.02, $\sigma$~= 0.05 \cite{jacobson11};
[Fe/H]~= $+$0.03, $\sigma$~= 0.07 \cite{overbeek15}.

\begin{figure}
\epsscale{0.90}
\plotone{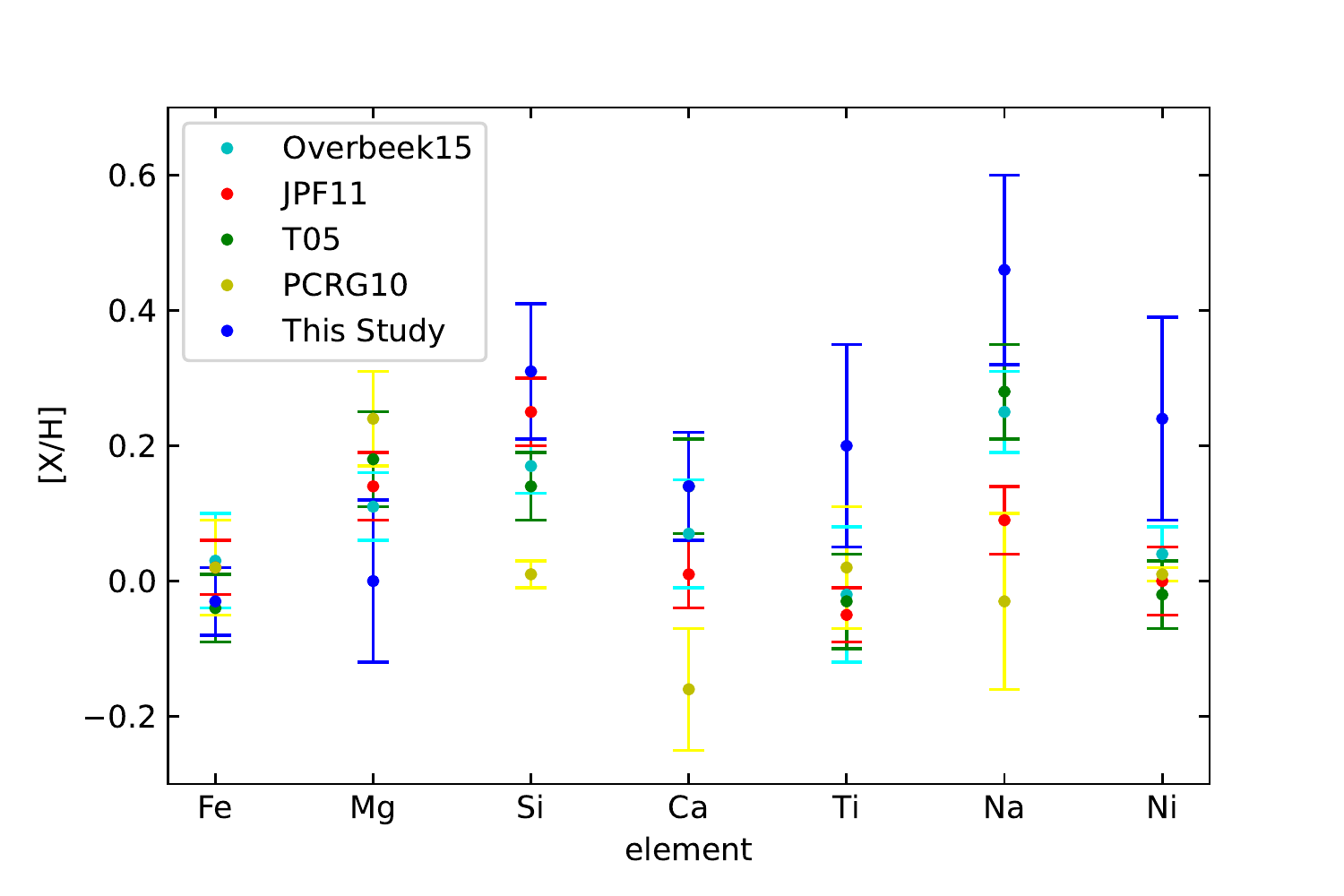}
\caption{\label{fig3}
\footnotesize
   Comparison of NGC 7789 abundances in this paper to results from some 
   previous papers.
   The various studies are identified by color types, as defined in the 
   plot legend and are as follows: \cite{overbeek15}, \cite{jacobson11}, \cite{pancino10}, \cite{tautvaisiene05}
}
\end{figure}

Our study is not a comprehensive abundance survey of NGC~7789.
It focuses mainly on the LiCNO element group and its relationship to
\species{He}{i} $\lambda$10830 absorption strengths.
We derived abundances from EWs of a few elements that have been featured in past
studies. 
They are listed in Table~\ref{tab-bigtable} and in Figure~\ref{fig3} we 
compare some [X/Fe] values to those in previous papers. 
Our abundance uncertainties are larger than those reported in the comparison
studies, but in most part they are in accord for Mg, Si, and Ca.
The NGC~7789 abundances for Ti, Na, and Ni should be explored further in a
study that concentrates on warmer, less line-rich stars than we have gathered
for our work.

\subsection{LITHIUM, CARBON, NITROGEN, AND OXYGEN}\label{cnolisyn}

\begin{figure}
\epsscale{0.70}
\center
\includegraphics[angle=-90,width=0.8\textwidth]{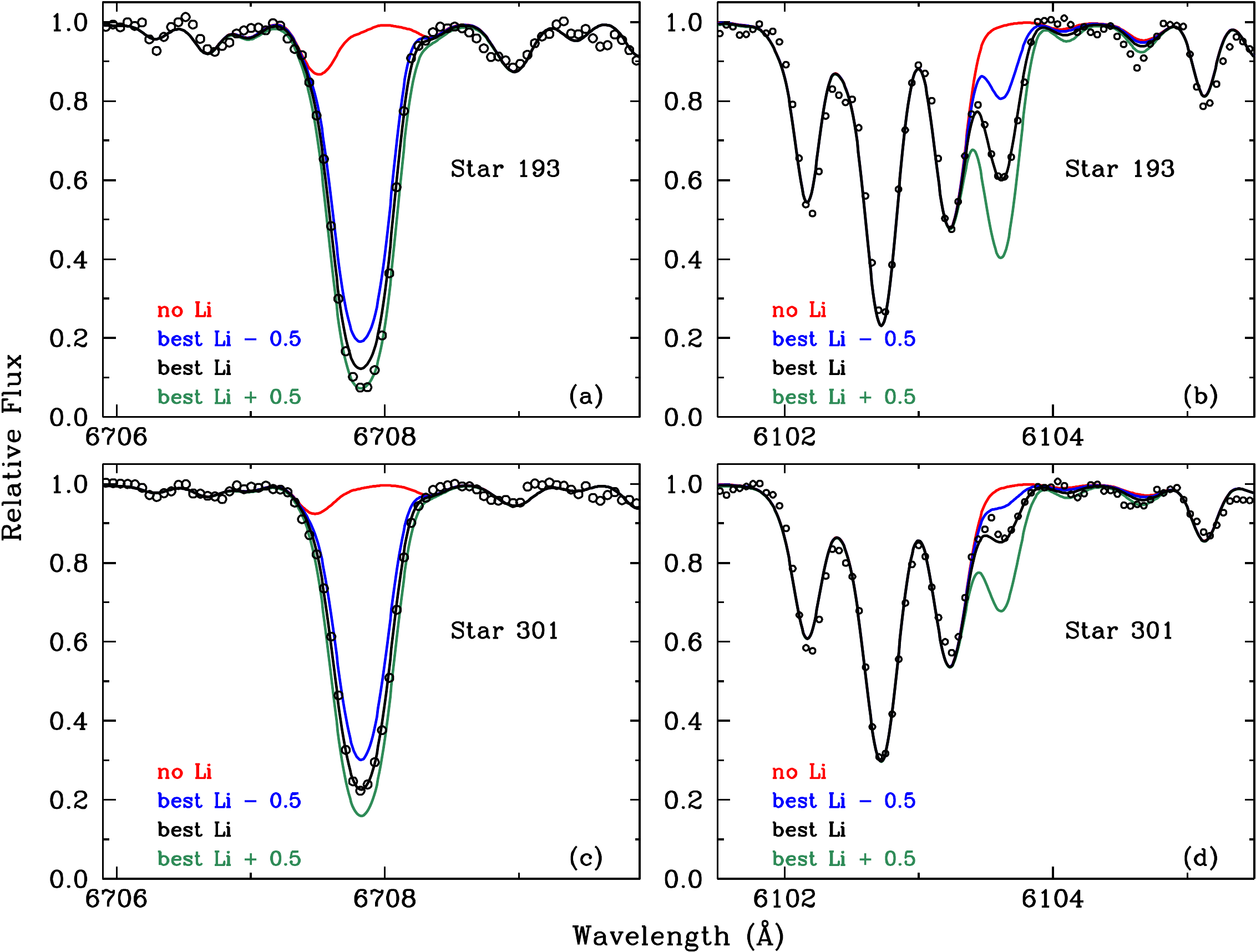}
\caption{\label{fig4}
\footnotesize
   Observed and synthetic spectra for the two NGC 7789 stars with very high
   Li abundances.
   The $\lambda$6707 resonance lines shown in panels (a) and (c) are so
   saturated that they cannot yield reliable abundances.
   The weaker, higher-excitation $\lambda$6103 lines in panels (b) and (d)
   are better Li features for these two stars.
   \citep{sneden22}. 
   In the legend of 
   each panel, "best'" indicates the synthesis using the abundance derived 
   for the named element, and "no'' indicates the synthesis computed without 
   any contribution of this element.  
   The other syntheses have abundance offsets (in dex) as indicated in 
   the legends.
}
\end{figure}

\begin{figure}
\epsscale{0.40}
\center
\includegraphics[angle=-90,width=0.8\textwidth]{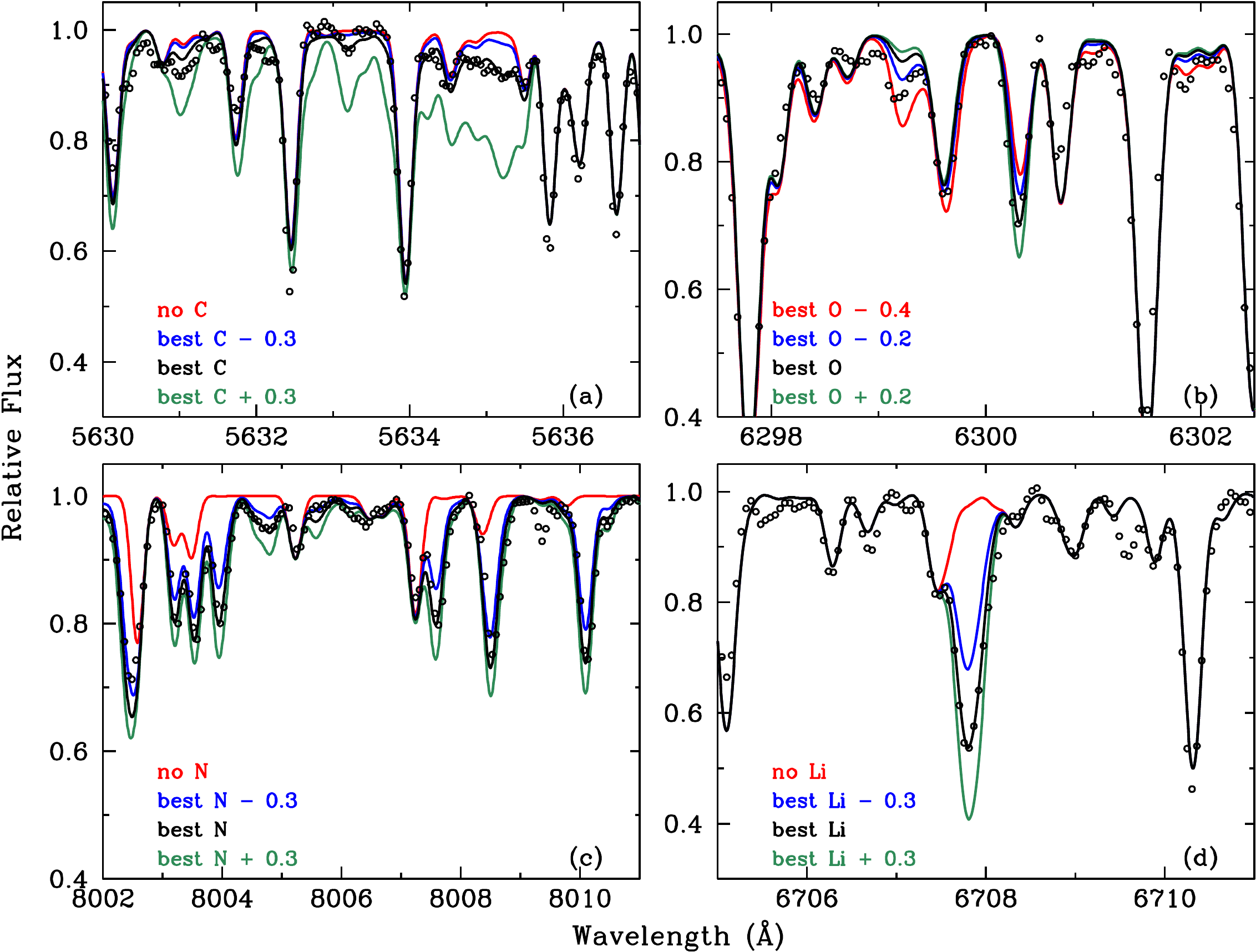}
\caption{\label{fig5}
\footnotesize
   Observed and synthetic spectra for the features used to determine C, N, 
   and O abundances in star 501, a typical program star.  
   The observed points are shown with open circles.  
   The meanings of the legends in each panel are the same as 
   in Figure~\ref{fig4}.
}
\end{figure}

Light elements Li, C, N, and O may be altered during a star's lifetime 
through interior proton fusion reactions and envelope mixing.
We derived abundances for these elements via matches of observed and
synthetic spectra.
\begin{itemize}
\item Lithium:  Nearly all Li abundances are derived from the 
\species{Li}{i} resonance doublet at 6707.8~\AA, but in a small percentage 
of cool stars with very large Li abundances the excited-state transition 
at 6103.6~\AA\ becomes detectable.  
We used both of these features in our analysis for stars 193 and 301.
\item Carbon:  the best available spectral features for our stars were
lines near the C$_2$ Swan system (d$^3\Pi_g-a^3\Pi_u$) (0$-$1) bandhead at 5635~\AA.
The stronger (0$-$0) bandhead near 5165~\AA\ is extremely crowded with
other atomic and molecular features, and our spectra in the region
of the comparable-strength (1$-$0) bandhead, 4725~\AA, are very noisy.
Our spectra are even less reliable  in the 4300~\AA\ 
CH G-band wavelength domain.  
\species{C}{i} high-excitation lines can be detected but are not reliable abundance indicators for NGC~7789 cool giants.  \item Nitrogen:  CN A$^2\Pi-X^2\Sigma^+$ red system lines can be detected  in all wavelength regions beyond 6000~\AA.  We concentrated on strong (2$-$0) band features near 8000~\AA\ to derive N abundances and \carbiso\ ratios.
\item The [\species{O}{i}] 
$\lambda$6300 ground-state
line is the single reliable
O abundance indicator in our spectra.
Care must be taken with this transition since it is beset with blending by
\species{Ni}{i} and CN contaminants, and can potentially be compromised
by telluric O$_2$ absorption and night-sky [\species{O}{i}] emission.
\end{itemize}

To create atomic/molecular line lists for the synthetic spectrum 
calculations we used
the $linemake$ facility \citep{placco21}\footnote{
https://github.com/vmplacco/linemake}.
The code creates lists of transitions starting with the 
\cite{kurucz11,kurucz18}\footnote{
http://kurucz.harvard.edu/linelists.html}
line compendium, and updating them with transition data from recent 
laboratory studies mainly by the Wisconsin-Madison atomic physics group
(\citealt{denhartog21} and references therein) and by the Old Dominion
University molecular physics group (\eg, \citealt{brooke16} and references
therein).

We estimated LiCNO abundances by comparing our observed spectra with synthetic
spectra computed with our model atmospheres (Table~\ref{tab-atmospheres}) and
these line lists.
Full molecular equilibrium calculations were performed as part of the
computations.
In Figure~\ref{fig4} we illustrate the observed/synthetic spectrum matches
for the \species{Li}{i} $\lambda$6707,6103 lines in the two Li-rich stars 
193 and 301.
Other program stars have Li abundances that are 2-3 orders of magnitude
smaller than those of stars 193 and 301, rendering weak $\lambda$6707 lines
(see also Figure~\ref{fig2}) and invisible $\lambda$6103 lines.

For C and O, the formation of CO molecules can significantly modify the C$_2$
and \species{O}{i} number densities, so we derived these abundances 
iteratively.
In Figure~\ref{fig5} we show the observed/synthetic matches for 
NGC 7789 star 501.
Note that although the C$_2$ band near 5635~\AA\ is always weak, it is very
sensitive to carbon abundance changes because of its double-carbon molecular
structure (see panel (a) of this figure).
Second, we repeat the caution from above that while the 
[\species{O}{i}]
transition at 6300.3~\AA\ is the only reliable oxygen abundance indicator
in optical spectra of red giant stars, it has significant blending by
other lines (panel (b) of the figure), and results from it should be treated 
with caution.
Finally, there are many \iso{12}CN features in the $\lambda$8000 region,
making total nitrogen abundance derivation straightforward.
However, in all of our NGC~7789 red giants the \iso{13}{CN} lines are much
weaker 
(Figure~\ref{fig5} panel c), and (like many other studies) we are forced to estimate 
\iso{12}C/\iso{13}C ratios mostly from the blended \iso{13}{CN} feature
at 8004.5~\AA.
Our derived LiCNO abundances are listed in Table~\ref{tab-helicno}.

\subsection{Abundance Uncertainties\label{uncertain}}

Our abundances
depend directly on uncertainties in 
EW
measurements and in synthetic/observed spectrum comparisons.
To assess EW uncertainties we repeated EW and synthetic spectrum
computations for multiple lines, varying line parameters and continuum 
placement.
We concluded that typical measurement uncertainties are $\pm$5\%,
contributing $\sim$ $\pm$0.03~dex to the abundances of individual
spectroscopic features.
The abundances also depend on the model atmosphere parameter choices.
Earlier in \S\ref{models}
we discussed the particular analytical
problem of NGC~7789 red giants.
All of our program stars have strong-lined spectra, and the majority of
our measured transitions are on or near the flat part of the curve of growth, 
where derived abundances become sensitive to assumed/derived microturbulent 
velocity \vmicro.
The difficulties
involved in using strong lines for model parameter
derivation should be kept in mind.
In Table~\ref{tab-depend} we list the responses of each species 
abundance to changes in model parameters that cover the range of expected 
uncertainties in \teff, logg, [M/H], and \vmicro.
For this table we have chosen star 765, whose parameters are roughly
in the middle of our stellar sample:  \teff~=~4515~K, \logg~=~2.3,
[M/H]~=~$-$0.1, \vmicro~=~1.4~\kmsec.
We present changes in abundances [X/H] 
(equivalent to changes in log~$\epsilon$)
to show how each species 
responds to
model parameter changes.
Lines of neutral species yield higher elemental abundances from \teff\ 
increases due to larger ionization, 
and abundances from lines
of ionized species
increase with increasing \logg\ due to decreased ionization in higher-gravity
stars.
Note that \species{V}{i} abundances make the largest changes as \teff\ varies;
this is why the LDR method  works very well for \teff\ estimation for
our cool giant stars.
Additionally, the Li abundance sensitivity to \teff\ and no other
atmospheric parameter in Table~\ref{tab-depend} confirms what has been
known in the literature about this species for decades.
Finally, in general the changes in [X/Fe] will be smaller than those of [X/H]
when comparisons of neutral to neutral and ion to ion are performed
as has been done in this paper. 

\section{DISCUSSION}\label{discussion}

\begin{figure}
\epsscale{0.70}
\center
\plotone{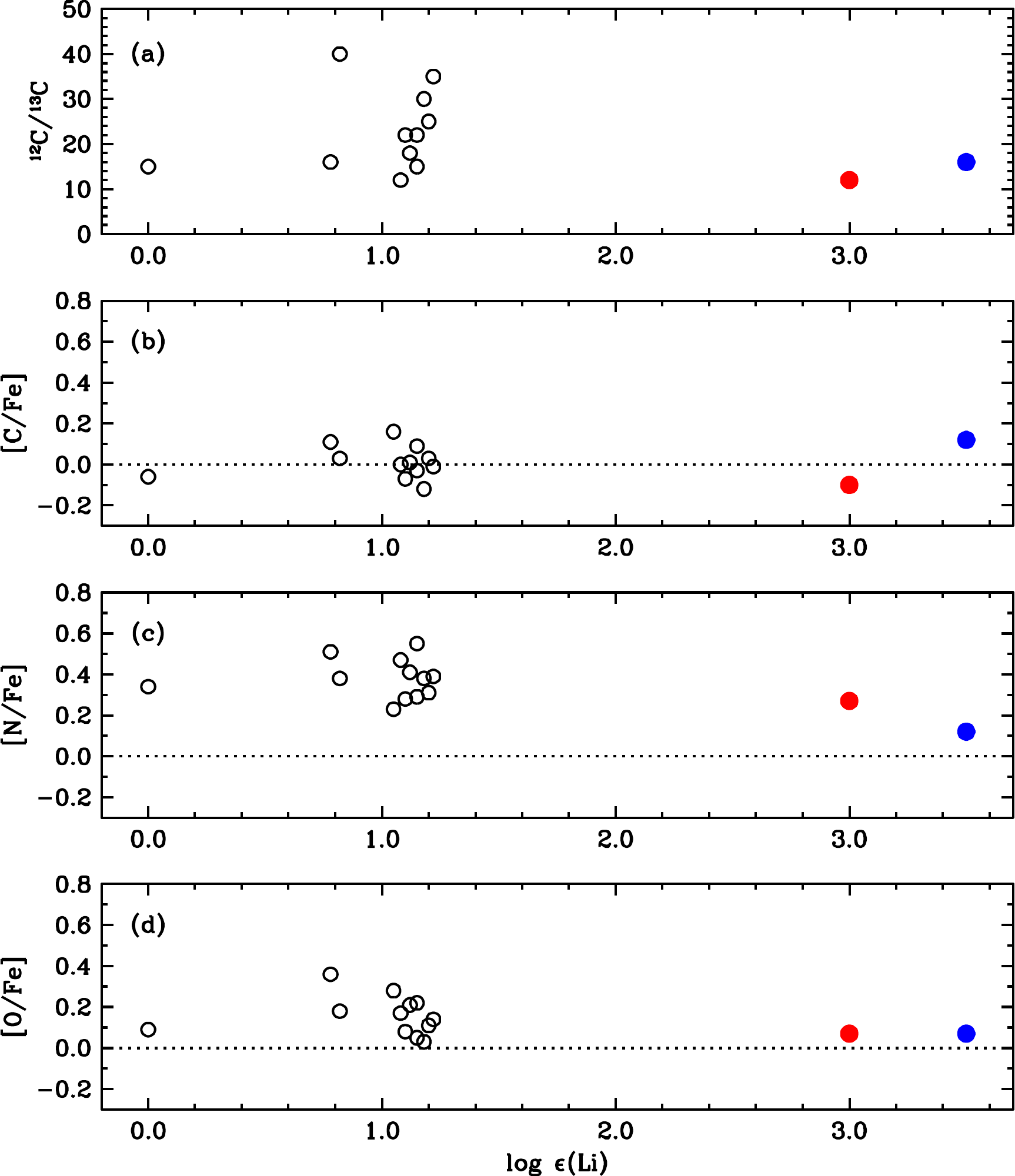}
\caption{\label{fig6}
\footnotesize
   Carbon isotopic ratios (panel a) and CNO abundances (panels b$-$d)
   as functions of Li abundance in all NGC~7789 program stars, including
   star 193, which is a probable non-member of the cluster.
   The dotted lines in panels (b$-$d) indicate the solar abundance 
   values.  
   Star 193 is the blue dot and Star 301 is the red dot.
}
\end{figure}

In this section, we discuss our main observational results on the light
elements in NGC~7789 red giants, and consider their relationships to
their probable evolutionary states.

\subsection{Li and CNO Abundances}\label{LiCNO}

These elements are participants in the various $p-p$ and CNO cycles
of hydrogen fusion, and we will consider them as a group.
In Figure~\ref{fig6} we plot the derived \carbiso\ and CNO
abundance ratios as functions of the Li abundances.
Inspection of this figure, as well as the mean values listed in 
Table~\ref{tab-helicno} suggest that there are few indications of significant
star-to-star CNO variations, and little evidence of their correlations
with Li.
For the CNO elements the abundance scatter is $\sigma$~$\simeq$~0.1, 
suggesting that all of our program stars have had similar abundance
changes in the first dredge-up near the base of the red giant branch (RGB).
Our derived C abundances in NGC~7789 do not appear to be much depleted,
$\langle$[C/Fe]$\rangle$~=~0.01.
In contrast, the classic studies 
of thin disk
population giants derived
lower C abundances.
For example, \cite{lambert81} found 
$\langle$[C/Fe]$\rangle$~=~$-$0.24$\pm$0.13.
We suspect that the difference stems from our exclusive use of the C$_2$
$\lambda$5635 bandhead for C abundances.
Our previous study of NGC~6940 \citep{bocek16} found that the $\lambda$5635
bandhead yields C abundances on average 0.17~dex higher than those derived
from the much stronger $\lambda$5165 bandhead.
As stated above, our spectra are not trustworthy in wavelength regions
less than about 5300~\AA.
However, if we were to shift our derived [C/Fe] abundance ratios downward by
$\simeq$0.15, then in consequence our CN-based N abundances would increase
by the same amount, from $\langle$[N/Fe]$\rangle$~=~0.35
(Table~\ref{tab-helicno}) to $\sim$0.50, compared with 
the \citeauthor{lambert81} value of $\langle$[N/Fe]$\rangle$~=~0.38$\pm$0.11.
Thus we regard our C and N abundances as 
probably
consistent with first dredge-up expectations.

Our carbon isotopic ratios are in the range \carbiso~=~12-40, 
with $\langle$\carbiso$\rangle$~=~21
($\sigma$~=~9), again consistent with values
that have been reported for field thin disk giants
for decades (\eg, \citealt{tomkin76} and references therein).

Most stars in our sample have similar Li abundances.  
From Table~\ref{tab-helicno}, excluding stars 
193, 301 and 765,
$\langle$\eps{Li}$\rangle$~=~1.08~$\pm$~0.04 ($\sigma$~0.15).
This value is shared by 11 of our 14 program stars. 
If stars 353 and 468 with slightly smaller Li abundances
(\eps{Li}~=~0.8) are excluded the average becomes more uniform but does not 
change much:  $\langle$\eps{Li}$\rangle$~=~1.14~$\pm$~0.02 ($\sigma$~0.05).
In total, the Li and CNO abundances of these stars probably result from
standard first dredge-up episodes that acted on a set of normal stars of
similar masses and initial chemical compositions.
Accurate mass measurements for individual 
stars
would be welcome but would 
require astroseismology. 

\begin{figure}
\epsscale{0.90}
\plotone{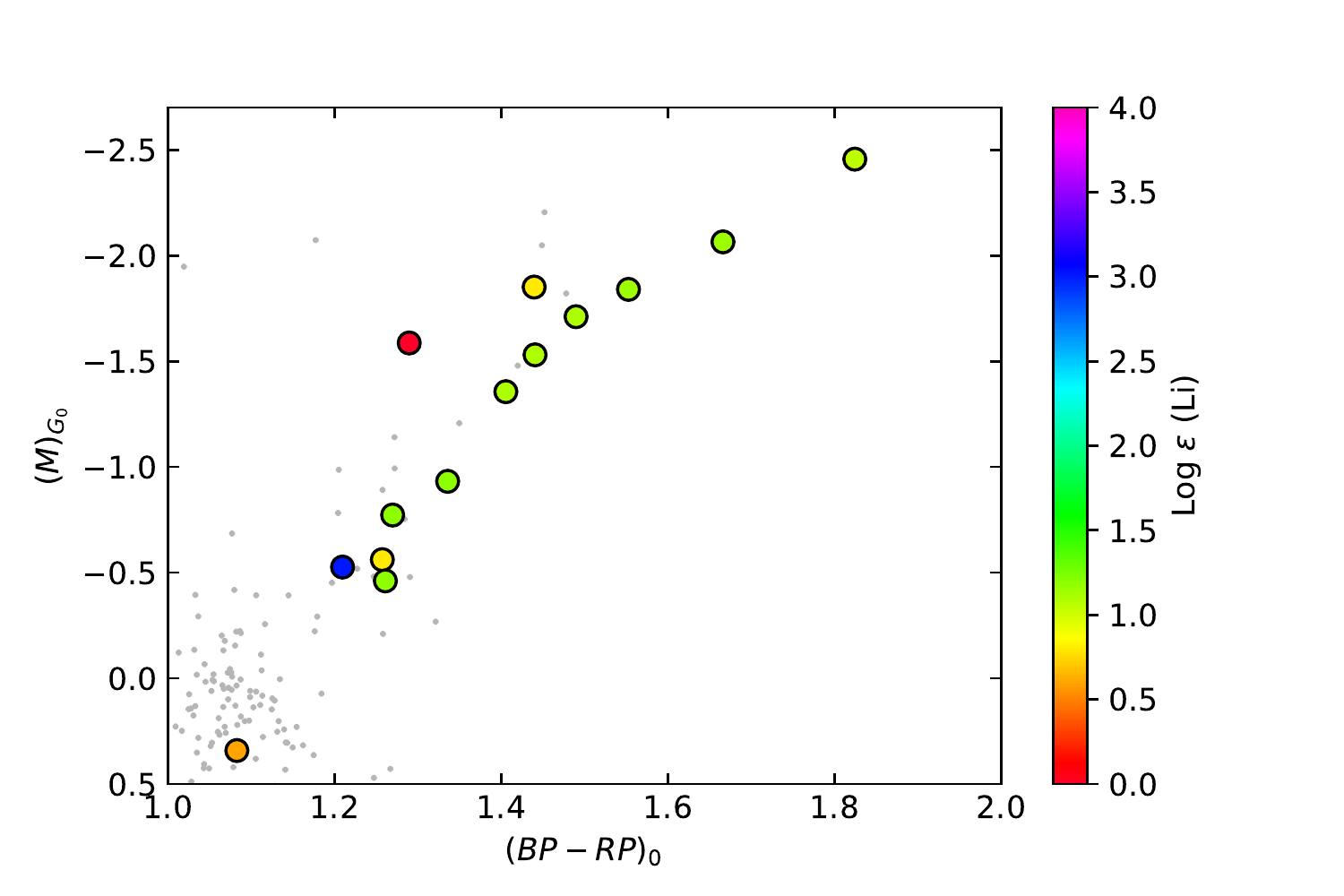}
\caption{\label{fig7}
\footnotesize
Another CMD for NGC 7789 with the data from Figure~\ref{fig1},
using colors to indicate the Li abundances for NGC~7789 members.
}
\end{figure}

To help clarify the evolutionary states of our stars, in 
Figure~\ref{fig7} we repeat Figure~\ref{fig1} except 
that different colors are used to 
indicate derived
lithium abundances.
For the present purpose we do not plot star 193, the probable 
non-member
of NGC~7789. 
We suggest that all stars with \eps{Li}~$\sim$~1 are first-ascent red giant
branch (RGB) members.
Star 468 
[$(BP-RP)_0$,$M_{G0}$] = [1.44,$-$1.85]
is slightly brighter 
than the first-ascent sequence and might be an asymptotic branch (AGB) star, 
but the separation is small.
A more convincing AGB assertion can be made for star 765 ([1.29,$-$1.59]), which is nearly a magnitude brighter than RGB
stars with similar colors, and its Li abundance \eps{Li}~=~0.0 is much smaller
than any other program star. Star 1101 ([1.08,0.34]) is our sole representative of the clump in NGC 7789, and being a
clump star means it has evolved beyond the first-ascent RGB.

Stars 193 (not plotted in Figure~\ref{fig7}) and 301 
(the blue point in the figure) may or may not require separate interpretations.
Although we agree with previous work suggesting that star 193 is not
a member of NGC~7789, that fact may have little impact on the interpretation
of its light element abundances.  
Stars 193 and 301 are solar-metallicity K giants with similar very high
Li abundances.
It is likely that both of them underwent similar evolutionary events that
resulted in their very high surface Li contents.
Star 193 departs from NGC~7789 cluster giants mainly in its relatively low 
[N/Fe] and consequently large [C/N] (Table~\ref{tab-helicno}).
Ejection of star 193 from NGC~7789 membership simply puts it in 
good company with the many field red giants of Paper~1 that have enhanced 
Li while showing no evidence for chromospheric activity anomalies.  

\subsection{COMPARISON WITH HELIUM}\label{Heanalysis}

\begin{figure}
\epsscale{0.90}
\plotone{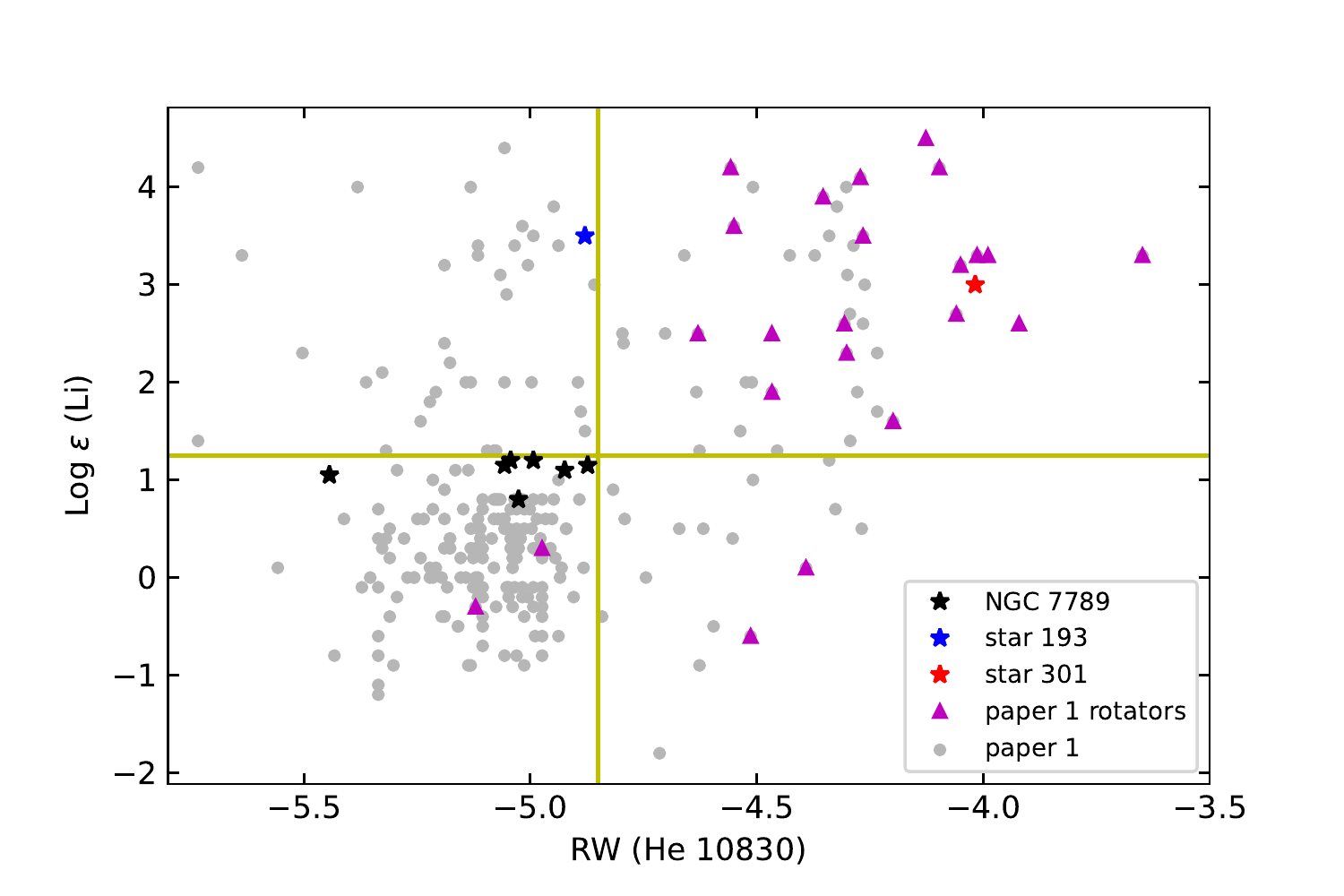}
\caption{\label{fig8}
\footnotesize
   Correlation between lithium abundances and He absorption line strengths
   in Paper 1 and the present work.
   Identification of the points are in the figure legend.
}
\end{figure}

Paper~1 reported \species{He}{i} $\lambda$10830 EWs for 10 NGC~7789 red
giants; see Table~\ref{tab-helicno}.
Five of the present program stars were not observed for Paper~1, and
star 225 was included in Paper~1 but not included in our optical survey.
In Figure~\ref{fig8} we plot 
Li abundances versus \species{He}{i} reduced widths
log(RW$_{\rm He}$).\footnote{
log(RW$_{\rm He}$) $\equiv$ log$_{10}$(EW$_{\rm He}/\lambda$) = 
log$_{10}$(EW$_{\rm He}$/10830)} 
We highlight the NGC~7789 stars and add in the field stars from Paper~1
for comparison.
Following the discussion in Paper~1, a yellow horizontal line is placed
at \eps{Li}~=~1.25 to approximately separate Li-rich giants from the majority 
of normal Li-poor stars.
A vertical line set at log(RW$_{\rm He}$)~=~$-$4.85 provides a suggested
separation between the vast majority of stars with weak \species{He}{i}
$\lambda$10830 absorption and the relatively few with anomalously strong
He lines.
These lines are rough guides and their values should not be interpreted
rigidly.

In Figure~\ref{fig8} all program stars with \eps{Li}~$\leq$~1.15 have weak 
$\lambda$10830 absorption lines, as do the vast majority of field red giants.
Probable non-member star 193 has a large Li abundance without excess He 
absorption, but many other field stars share this characteristic.
Member star 301 has both a high Li and a large $\lambda$10830 feature.  
In fact, this star has one of the largest \species{He}{i} lines found in the
Paper~1 survey.
All other stars with log(RW$_{\rm He}$)~$\gtrsim$~$-$4.2 exhibit significant
rotation spectral line broadening, with derived velocities in the range
\vsini~=~9$-$140~\kmsec.
Paper~1 did not report rotational line broadening for star 301 but the derived
Gaussian smoothing was larger than for most stars. 
Therefore we have carefully re-examined its $\lambda$10830 region with 
synthetic spectra.
We now tentatively can assign a small rotational value \vsini~=~6$\pm$1~\kmsec\
for star 301.
However, this velocity broadening is nearly at the limit of our ability
to isolate rotation from the other broadening sources (thermal, microturbulent,
macrotrubulent, and instrumental), so caution should be exercised in 
interpretation of this value.
What is not in dispute is the relatively small total line broadening of
star 301 in comparison to many other Li-rich red giants.

The vast majority of the 300-star field red giant sample in Paper~1 
proved to be mostly stars with colors and magnitudes consistent with the 
red clump and general red horizontal branch.
Stars with large Li abundances and \species{He}{i} $\lambda$10830 line
strengths both weak and strong were as likely to occupy this CMD domain
as those with neither Li nor He anomalies. 
The interpretation in Paper~1 centered on assignment of fresh Li production 
to the helium flash that preceded appearance of the stars on the red 
horizontal branch, as outlined in \S\ref{intro} (see also \citealt{mallick23}).
This idea may be applicable for stars 301 and 193 as well.
Star 301 has high Li and its strong $\lambda$10830 line argue for recent 
helium-flash disturbance of its outer envelope.

However, star 301's evolutionary state is still uncertain.
Based on its CMD position, certain possibilities arise.
Star 301 could be ascending the RGB, in which case it would not have undergone 
any helium flash yet, and so both Li and He features would likely be due to 
interactions with a binary companion. 
We did not detect any binary companion in star 301's spectrum.
In this case, this star's apparent excess rotation suggests that it could 
have an unseen companion like a white dwarf or small main sequence star that 
is simply too dim to contribute any significant flux or spectral features. 
Figure 8 of \cite{casey19} presents three models for how the presence of a 
binary companion(s) like star(s) or planet(s) can explain the lithium-richness 
of certain red giant stars.
If star 301 is still ascending the RGB then tidal interactions with the 
putative companion are to blame.
A radial velocity monitoring program might be useful for detecting the  
companion.
More directly, speculation on star 301's ``late'' arrival on the RGB probably
requires it to have recently evolved from its previous status as a cluster BSS.
But such a scenario must account for star 301's carbon istopic ratio, which 
is among the smallest of any star in our sample (Table~\ref{tab-helicno}).
Blame for its small \carbiso\ would shift to the companion's transferred
material.

Star 301 might be a normal red clump giant, even though its absolute 
magnitude ($M_{G0}$) appears to be $\sim$0.5~mag brighter than the 
NGC~7789 clump (Figure~\ref{fig7}).
No reasonable amount of (unproven) differential reddening corrections 
would place star 301 in the NGC~7789 clump (see the reddening vector in 
Figure~\ref{fig1}).
Finally, NGC 7789 may have a more complex member population than most 
open clusters. 
From investigation of CN and CH molecular band strengths of NGC~7789 giant
stars, \cite{carrera13} suggested that this cluster may have a CN intracluster
spread of the sort seen in typical globular clusters.  
They urged caution in interpretation of their observations
pending a much larger CN survey, but it is possible the star 301 is a 
representative of an anomalous sub-population of NGC~7789.
If this is true, then star 301 should be dealt with as a special evolutionary
case. 
Asteroseismological data for this star would help its interpretation.

Star 193 has an absolute magnitude more consistent with a red clump star
but it appears to be too red (Figure~\ref{fig1}).
This could be alleviated by assignments of a larger reddening value for
this star than what we have assumed for the cluster. 
Though star 193 is a helium
flash candidate, its lithium could too be explained by a binary companion as one of the
binary companion models in \cite{casey19} involve lithium richness in post-he-flash stars but due to a binary 
companion rather than the helium flash.

Further progress in this area requires increasing the number of
Li-rich red giants that have \species{He}{i} $\lambda$10830 data and other
common characteristics.
Efforts by our group are underway to gather such observations for
Kepler field giants, whose evolutionary states can be clearly defined
through asteroseismology.
We are also collecting spectra of red giants with known rotational velocities
to test that possible connection.
Additionally the number of Li-rich red giant members in other open clusters
is slowly growing (\citealt{magrini21} and references therein).
Observational campaigns on these clusters should be undertaken in the
future.

\section{CONCLUSIONS}\label{conclusions}

In this paper we have described the derivation of model atmospheric 
parameters and LiCNO abundances for 15 red giants of the open cluster NGC~7789.
We have explored the relationship between Li abundances, CNO abundances, and He $\lambda$10830
absorption strengths.
We find that the majority of observed NGC 7789 stars have consistent 
Li abundances, \eps{Li}~$\simeq$~ 1.0.
Two anomalous stars 193 and 301 have \eps{Li} of 3.5 and 3.0 
respectively; they are unambiguous examples of the Li-rich red giant subclass.
The probable non-membership of star 193 hinders a detailed investigation of
its evolutionary state, but its Li abundance is well determined.
The \species{He}{i} $\lambda$10830 transition does not lead to an easy
interpretation of the Li richness of stars 193 and 301, given the feature 
is present in one star but not the other.
In concert with other papers on Li abundance enhancements in field red giants,
we suggest (but cannot prove) that the Li and He peculiarities detected
in a small percentage of giants are due to recent or long-ago envelope
turbulence that accompanied the helium flashes in these stars.
Li abundances and He line strength measurements for many other
NGC~7789 red giants should go a long way to clarifying the overall 
stellar evolution story of this cluster.

\begin{acknowledgments}
We thank the referee for helpful comments that improved the paper
These results are based on observations obtained with the Tull 
Spectrograph aboard the 2.7 meter Harlan J Smith Telescope at McDonald 
Observatory. 
We acknowledge support from NSF grant AST-1616040 (CS). 
We also appreciate additional financial support and resources from the 
University of Texas at Austin's Astronomy Department and the College of 
Natural Sciences.
This work has made use of data from the European Space Agency (ESA) 
mission {\it Gaia} (\url{https://www.cosmos.esa.int/gaia}), processed by the 
{\it Gaia} Data Processing and Analysis Consortium (DPAC,
\url{https://www.cosmos.esa.int/web/gaia/dpac/consortium}). 
Funding for the DPAC has been provided by national institutions, in particular 
the institutions participating in the {\it Gaia} Multilateral Agreement.
\end{acknowledgments}

\facility{Smith (Tull Spectrograph), HET (HPF)}

\software{linemake (https://github.com/vmplacco/linemake), 
MOOG (Sneden 1973), 
IRAF (Tody 1986, Tody 1993), 
SPECTRE (Fitzpatrick \& Sneden 1987)}

\bibliographystyle{aasjournal}



\clearpage
\begin{center}
\begin{deluxetable}{cc}
\tablewidth{0pt}
\tablecaption{Basic Data for NGC 7789\label{tab-cluster}}
\tablecolumns{2}
\tablehead{
\colhead{Quantity}                                &
\colhead{Value\tablenotemark{a}}
}
\startdata
RA                               &  23:57:24                      \\
Dec                             &  +56:42:30                    \\
$l$ (Galactic)              &  115.53$^{\circ}$           \\
$b$ (Galactic)             &  $-$5.53$^{\circ}$         \\
Parallax                       &  0.45 mas                     \\
Distance                      &  2337 pc                       \\
Reddening                  &  0.28 mag                      \\
Distance Modulus       &  11.72                            \\
Age                             &  1.5 Gyr                       \\
$\mu$ (RA)                 &  $-$0.92 mas                 \\
$\mu$ (Dec)                &  $-$1.93 mas                \\
Radial Velocity            &  $-$54.5 \kmsec            \\
\enddata

\tablenotetext{a}{from WEBDA or SIMBAD except in the case of the Reddening and the age (age is from \citep{gao18}). For the reddening, we used the value found in \citep{jacobson11}\ and \citep{overbeek15}\ quoting \citep{gim98}\, and \citep{tautvaisiene05}\ }

\end{deluxetable}                                                   
\end{center}  

\begin{center}
\begin{deluxetable}{crrrrrrr}
\tablewidth{0pt}
\tablecaption{Basic Data for Target Stars\tablenotemark{a}\label{tab-stars}}
\tablecolumns{8}
\tablehead{
\colhead{Star\tablenotemark{b}}       		&       
\colhead{parallax}                  		&
\colhead{$\mu$ (RA)}                        &
\colhead{$\mu$ (Dec)}                       &
\colhead{$G$}                               &
\colhead{$BP-RP$}                           &
\colhead{$V$}                               &
\colhead{$B-V$}                             \\
\colhead{}                                  &
\colhead{mas}                               &
\colhead{mas}                               &
\colhead{mas}                               &
\colhead{}                                  &
\colhead{}                                  &
\colhead{}                                  &
\colhead{}                                 
}
\startdata
72 	  &	0.5040  & $-$0.775 	& $-$2.071 & 10.22	& 2.06 &  11.05 & 1.74	 \\
193	  &	0.6742  & $+$8.491	& $-$2.338 & 12.12	& 1.63 &  12.61 & 1.42    \\
301	  &	0.5144	& $-$1.123	& $-$2.444 & 11.72	& 1.60 &  12.27 & 1.34	 \\
329	  &	0.4949	& $-$0.883	& $-$1.890 & 11.80	& 1.66 &  12.31 & 1.38    \\
353	  &	0.4855	& $-$0.820	& $-$2.017 & 12.10	& 1.65 &  12.59 & \nodata \\
461	  &	0.4793	& $-$0.968	& $-$1.933 & 10.69	& 1.88 &  11.35 & 1.63	 \\
468	  &	0.4829	& $-$0.886	& $-$1.914 & 10.53	& 1.83 &  11.61 & 1.14	 \\
494	  &	0.4898	& $-$0.825	& $-$1.985 &  9.86	& 2.22 &  10.74 & 1.68	 \\
501	  &	0.4749	& $-$0.830	& $-$2.037 & 10.51	& 1.95 &  11.26 & 1.71	 \\
637	  &	0.4882	& $-$0.983	& $-$1.716 & 11.90	& 1.65 &  12.40 & 1.42    \\
669	  &	0.4830	& $-$0.775	& $-$2.071 & 10.85	& 1.83 &  11.46 & 1.58	 \\
765	  &	0.4175	& $-$1.065	& $-$2.033 & 11.11	& 1.68 &  11.59 & 1.21	 \\
970	  &	0.4800  & $-$0.970	& $-$2.110 & 11.29	& 1.80 &  11.89 & \nodata \\
971   &	0.5398	& $-$1.029	& $-$2.105 &  9.97	&      &  11.05 & 1.90	 \\
1066  &	0.4867	& $-$0.979	& $-$2.055 & 11.43	& 1.73 &  11.99 & \nodata \\
1101  &	0.5004	& $-$0.876	& $-$2.013 & 12.65	& 1.48 &  13.05 & \nodata \\
      &         &           &          &        &      &        &         \\
$<>$\tablenotemark{c}
      & 0.4881  & $-$0.919  & $-$2.046 &        &      &        &         \\
$\sigma$\tablenotemark{c}
      & 0.0256  &    0.107  &    0.154 &        &      &        &         \\
\enddata

\tablenotetext{a}{adopted from SIMBAD}
\tablenotetext{b}{\cite{kustner23}}
\tablenotetext{c}{computed without star 193}

\end{deluxetable}                                                   
\end{center}

\begin{center}
\begin{deluxetable}{crrrrrrrrrrrrrrrrrr}
\tablewidth{0pt}
\tablecaption{Equivalent Widths\label{tab-ews}}
\tablecolumns{19}
\tabletypesize{\small}
\tablehead{
\colhead{Species}                           & 
\colhead{$\lambda$}                         & 
\colhead{$\chi$}                            & 
\colhead{log($gf$)}                         & 
\colhead{72}                                & 
\colhead{193}                               & 
\colhead{301}                               & 
\colhead{329}                               & 
\colhead{353}                               & 
\colhead{461}                               & 
\colhead{468}                               & 
\colhead{494}                               & 
\colhead{501}                               & 
\colhead{637}                               & 
\colhead{669}                               & 
\colhead{765}                               & 
\colhead{978}                               & 
\colhead{1066}                              & 
\colhead{1101}                               
}
\startdata 
Na I & 5682.64& 2.101& -0.70& 216& 194& 173& 181& 175& 210& 209& 212& 217& 186& 197& \nodata&
                              194& 180& 146\\
Na I & 6154.23& 2.101& -1.56& 139& 109&  94&  98&  97& 128& 138& 144& 133&  99& 118& 105& 112& 105&  71\\
Na I & 6160.75& 2.103& -1.26& 154& 130& 112& 120& 114& 142& 144& 195& 147& 118& 136& 121& 132& 125&  93\\
Mg I & 5528.41& 4.343& -0.62& 290& 321& 275& 266& 261& 299& 290& 291& \nodata& 
                              236& 285& \nodata& 272& 279& 209\\
Mg I & 5711.08& 4.343& -1.83& 154& 166& 143& 144& 146& 161& 141& 148& 157& 152& 161& 154& 151& 151& 113\\
Mg I & 7811.11& 5.941& -0.95&  99& 143&  80&  79&  81&  94& \nodata& 
                              101&  93&  86&  91&  85&  81&  92& \nodata\\
Al I & 6696.02& 3.140& -1.35& 140& 212&  99& 131& 123& 163& 121& 150& 126& 102& 148& 101& 109& 130&  72\\
Al I & 6698.67& 3.140& -1.64&  90&  88&  54&  76&  66&  91&  88&  99&  88&  74&  86&  73&  78&  75&  41\\
Al I & 7835.30& 4.018& -0.65& 111& 125&  85&  81&  74& 103&  93& 117& 109&  79&  98&  81&  97&  85&  64\\
Si I & 5488.98& 5.614& -1.90&  29&  54&  33&  44&  42&  33&  36&  25&  31&  46&  40&  31&  41&  41&  34\\
Si I & 5517.53& 5.082& -2.61&  22&  35&  33&  27&  23&  32& \nodata&  
                               35& \nodata&  36&  34&  31& \nodata&  28&  25\\
\enddata

\tablecomments{All EWs are in units of \AA\ (This table is available in its 
entirety in machine-readable form.)}

\end{deluxetable}
\end{center}

\begin{center}
\begin{deluxetable}{rrrrrrrrrrr}
\tablewidth{0pt}
\tablecaption{Model atmosphere parameters\label{tab-atmospheres}}
\tablecolumns{11}
\tablehead{
\colhead{Star}              &       
\colhead{\teff}               &
\colhead{[M/H]}            &
\colhead{\logg}             &
\colhead{\vmicro}         &
\colhead{[Fe/H]}           &
\colhead{$\sigma$}      &
\colhead{\#lines}          &
\colhead{[Fe/H]}           &
\colhead{$\sigma$}      &
\colhead{\#lines}          \\
\colhead{}                    &
\colhead{(K)}               &
\colhead{}                    &
\colhead{}                    &
\colhead{\kmsec}        &
\colhead{(I)}                &
\colhead{(I)}                &
\colhead{(I)}                &
\colhead{(II)}               &
\colhead{(II)}               &
\colhead{(II)}              
}
\startdata
72     & 4240 & $-$0.1 & 2.00 & 1.00 &    0.01 & 0.18 & 50 &    0.01 & 0.26 & 12 \\ 
193    & 4590 & $-$0.1 & 2.50 & 1.60 & $-$0.02 & 0.10 & 49 & $-$0.02 & 0.16 & 12 \\
301    & 4730 & $-$0.1 & 2.60 & 1.55 & $-$0.06 & 0.13 & 49 & $-$0.07 & 0.17 & 11 \\	
329    & 4625 & $-$0.1 & 2.40 & 1.40 &    0.04 & 0.12 & 52 & $-$0.03 & 0.17 & 12 \\
353    & 4675 & $-$0.1 & 2.40 & 1.60 & $-$0.13 & 0.12 & 50 & $-$0.13 & 0.22 & 10 \\
461    & 4315 & $-$0.1 & 2.10 & 1.50 & $-$0.01 & 0.15 & 48 & $-$0.01 & 0.14 &   7 \\
468    & 4370 & $-$0.1 & 2.10 & 1.60 & $-$0.05 & 0.14 & 44 & $-$0.07 & 0.14 & 10 \\
494    & 4170 & $-$0.1 & 1.90 & 1.45 &    0.02 & 0.25\tablenotemark{a} & 
                                                        49 &    0.02 & 0.19 &   9 \\
501    & 4295 & $-$0.1 & 2.10 & 1.50 &    0.00 & 0.15 & 46 &    0.00 & 0.11 &   7 \\
637    & 4620 & $-$0.1 & 2.60 & 1.50 & $-$0.06 & 0.10 & 51 & $-$0.06 & 0.08 &   9 \\
669    & 4390 & $-$0.1 & 2.00 & 1.45 & $-$0.03 & 0.14 & 50 & $-$0.02 & 0.33 & 10 \\
765    & 4515 & $-$0.1 & 2.30 & 1.40 &    0.07 & 0.15 & 50 &    0.06 & 0.12 & 10 \\
970    & 4455 & $-$0.1 & 2.30 & 1.55 & $-$0.02 & 0.13 & 49 & $-$0.02 & 0.08 &   9 \\
1066   & 4530 & $-$0.1 & 2.30 & 1.50 & $-$0.03 & 0.13 & 53 & $-$0.02 & 0.15 &   9 \\
1101   & 5050 & $-$0.1 & 2.65 & 1.40 & $-$0.08 & 0.12 & 47 & $-$0.08 & 0.13 &   7 \\
       &      &        &      &      &         &      &    &         &      &     \\
$<>$   &      &        &      &      & $-$0.02 & 0.14 & 49 & $-$0.03 & 0.16 &  10\\
$\sigma$&  225  &\nodata& 0.24  & 0.07 &     0.05 & 0.04 &\nodata &0.05 & 0.07 &\nodata \\
\enddata

\tablenotetext{a}{Star 494 had an exceptionally low signal to noise (SNR) which is why we ended up with an unusually large standard deviation. In addition, the reason there is such a wide range of standard deviations for Fe (II) is that many Fe (II) lines lay in noisy regions of our spectra, making them harder to measure and depending on the star.}

\end{deluxetable}                                                   
\end{center}  

\begin{center}
\begin{deluxetable}{rrrrrrrrrrrrrrrrrr}
\tabletypesize{\small}
\tablewidth{0pt}
\tablecaption{Output Abundances in bracket ([X/Fe]) form\label{tab-bigtable}}
\tablecolumns{18}
\tablehead{
\colhead{Star}                         		&       
\colhead{Na}			    		&
\colhead{Mg}				    	&
\colhead{Al}					&
\colhead{Si}					&
\colhead{Ca}					&
\colhead{Sc2\tablenotemark{a}}		&
\colhead{Ti1}					&
\colhead{Ti2}					&
\colhead{V\tablenotemark{a}}		&
\colhead{Cr1}					&
\colhead{Cr2}					&
\colhead{Mn\tablenotemark{a}}		&
\colhead{Fe1}					&
\colhead{Fe2}					&
\colhead{Ni}					&
\colhead{La\tablenotemark{a}}		&
\colhead{Eu\tablenotemark{a}}		
}
\startdata
72    & 0.61 &      0.01 & 0.33 & 0.29 & 0.24 &      0.48 & 0.44 &      0.51 & 0.49 & 0.31 & 0.26       &      0.29 & 0.00 & 0.00 & 0.47  & 0.35 & 0.22 \\
193  & 0.46 &      0.27 & 0.83 & 0.50 & 0.17 &      0.15 & 0.20 &      0.23 & 0.22 & 0.18 & 0.32      &      0.20 & 0.00 & 0.00 & 0.24   & 0.18 & 0.55 \\
301  & 0.38 & $-$0.03 & 0.12 & 0.20 & 0.32 &      0.13 & 0.20 &      0.33 & 0.20 & 0.31 & 0.23      &      0.05 & 0.00 & 0.00 & 0.12   & 0.24 & 0.17 \\
329  & 0.45 & $-$0.04 & 0.35 & 0.35 & 0.17 &      0.09 & 0.14 &      0.33 & 0.14 & 0.18 & 0.15      &      0.13 & 0.00 & 0.00 & 0.22   & 0.25 & 0.28 \\
353  & 0.45 &      0.03 & 0.29 & 0.38 & 0.16 &      0.10 & 0.15 &      0.22 & 0.21 & 0.18 & 0.20      &      0.11 & 0.00 & 0.00 & 0.15   & 0.29 & 0.32 \\
461  & 0.50 &      0.04 & 0.43 & 0.39 & 0.18 &      0.20 & 0.27 &      0.53 & 0.34 & 0.18 & 0.29      &      0.22 & 0.00 & 0.00 & 0.33   & 0.27 & 0.29 \\
468  & 0.64 & $-$0.13 & 0.22 & 0.42 & 0.20 &      0.16 & 0.28 &      0.16 & 0.32 & 0.15 & 0.07      &      0.21 & 0.00 & 0.00 & 0.35   & 0.37 & 0.48 \\
494  & 0.77 &      0.01 & 0.43 & 0.39 & 0.18 &      0.11 & 0.47 &      0.69 & 0.58 & 0.30 & 0.53       &      0.12 & 0.00 & 0.00 & 0.46  & 0.36 & 0.19 \\
501  & 0.57 &      0.19 & 0.26 & 0.29 & 0.18 &      0.16 & 0.33 &      0.41 & 0.41 & 0.33 & 0.20      &      0.13 & 0.00 & 0.00 & 0.35   & 0.38 & 0.46 \\
637  & 0.41 & $-$0.04 & 0.16 & 0.42 & 0.12 &      0.20 & 0.13 &      0.26 & 0.19 & 0.15 & 0.37      &      0.16 & 0.00 & 0.00 & 0.25   & 0.35 & 0.50 \\
669  & 0.52 &      0.07 & 0.41 & 0.28 & 0.20 &      0.16 & 0.32 &      0.42 & 0.32 & 0.23 & 0.31      &      0.30 & 0.00 & 0.00 & 0.31   & 0.24 & 0.27 \\
765  & 0.27 &      0.10 & 0.01 & 0.24 & 0.09 &      0.14 & 0.13 &      0.19 & 0.13 & 0.12 & 0.07      &      0.17 & 0.00 & 0.00 & 0.25   & 0.21 & 0.36 \\
970  & 0.44 & $-$0.07 & 0.16 & 0.40 & 0.12 &      0.13 & 0.14 &      0.29 & 0.21 & 0.10 & 0.17      &      0.15 & 0.00 & 0.00 & 0.25   & 0.31 & 0.35 \\
1066& 0.41 &      0.02 & 0.27 & 0.38 & 0.19 &      0.10 & 0.13 &      0.45 & 0.15 & 0.15 & 0.12      &      0.00 & 0.00 & 0.00 & 0.24   & 0.25 & 0.26 \\
1101& 0.43 &      0.03 & 0.06 & 0.19 & 0.05 & $-$0.07 & 0.07 & $-$0.03 & 0.04 & 0.02 & $-$0.21 & $-$0.07 & 0.00 & 0.00 & 0.05   & 0.21 & 0.45 \\
\enddata

\tablenotetext{a}{For these 5 elements, the abundances presented were found using the "blends" task instead of "abfind" due to the presence of hyperfine splitting. All other abundances were found using "abfind".}

\end{deluxetable}                                                   
\end{center}  

\begin{center}
\begin{deluxetable}{rrrrrrrrr}
\tabletypesize{\small}
\tablewidth{0pt}
\tablecaption{LICNO Abundances, Carbon Isotopic Ratios, He EWs}\label{tab-helicno}
\tablecolumns{9}
\tablehead{
\colhead{Star}                    &       
\colhead{\iso{12}{C}/\iso{13}{C}} &
\colhead{[Fe/H]}				  &
\colhead{[C/Fe]}				  &
\colhead{[N/Fe]}				  &
\colhead{[O/Fe]}				  &
\colhead{[C/N]}		              &
\colhead{\eps{Li}}                &
\colhead{EW(He)\tablenotemark{a}}
}
\startdata
72       & 15	   &    0.01 &    0.09 & 0.29  & 0.22  & $-$0.20 & 1.15  &   95    \\
193      & 16	   & $-$0.02 &    0.12 & 0.12  & 0.07  &    0.00 & 3.50  &  143    \\
301      & 12      & $-$0.07 & $-$0.10 & 0.27  & 0.07  & $-$0.37 & 3.00  & 1040    \\	
225      & \nodata & \nodata & \nodata &\nodata&\nodata& \nodata &\nodata&  102    \\
329      & 35      & $-$0.04 & $-$0.01 & 0.39  & 0.14  & $-$0.40 & 1.20  &   98    \\
353      & 40      & $-$0.13 &    0.03 & 0.38  & 0.18  & $-$0.35 & 0.80  &  102    \\ 
461      & 18	   & $-$0.01 &    0.01 & 0.41  & 0.21  & $-$0.40 & 1.10  &  129    \\
468      & 16      & $-$0.06 &    0.11 & 0.51  & 0.36  & $-$0.40 & 0.80  & \nodata \\
494      & \nodata &    0.02 &    0.16 & 0.23  & 0.28  & $-$0.07 & 1.05  &   39    \\
501      & 22      &    0.00 & $-$0.03 & 0.55  & 0.05  & $-$0.58 & 1.15  &  145    \\
637      & 25      & $-$0.06 &    0.03 & 0.31  & 0.11  & $-$0.28 & 1.20  &  110    \\
669      & 22      & $-$0.03 & $-$0.07 & 0.28  & 0.08  & $-$0.35 & 1.10  & \nodata \\
765      & 15      &    0.06 & $-$0.06 & 0.34  & 0.09  & $-$0.40 & $\lesssim$0.0   & \nodata \\
970      & 12      & $-$0.02 &    0.00 & 0.47  & 0.17  & $-$0.47 & 1.10  & \nodata \\
1066     & 30	   & $-$0.03 & $-$0.12 & 0.38  & 0.03  & $-$0.50 & 1.20  & \nodata \\
1101     & 15	   & $-$0.08 &    0.08 & 0.33  & 0.08  & $-$0.25 & $\lesssim$0.5   & \nodata \\
         &         &         &         &       &       &         &       &         \\
$<>$     & 21      & $-$0.03 &    0.02 & 0.35  & 0.14  & $-$0.33 &       &         \\
$\sigma$ &  9      &    0.05 &    0.10 & 0.11  & 0.10  & $-$0.16 &       &         \\
\enddata

\tablenotetext{a}{Taken from Paper 1}

\end{deluxetable}                                                   
\end{center}

\begin{center}
\begin{deluxetable}{rrrrr}
\tabletypesize{\small}
\tablewidth{0pt}
\tablecaption{Abundance Dependencies on Model Atmosphere Parameters\label{tab-depend}}
\tablecolumns{5}
\tablehead{
\colhead{Quantity}                 &
\colhead{$\Delta$(log $\epsilon$)} &
\colhead{$\Delta$(log $\epsilon$)} &
\colhead{$\Delta$(log $\epsilon$)} &
\colhead{$\Delta$(log $\epsilon$)}
}
\startdata
parameter            & $\Delta$\teff\ & $\Delta$\logg\ & $\Delta$[M/H]
                     & $\Delta$\vmicro\ \\
change               & $+$100 K  &   $+$0.2  &   $+$0.3  & $-$0.2\kmsec \\
\species{Fe}{i}      &     0.02  &     0.02  &     0.06  &     0.10 \\
\species{Fe}{ii}     &  $-$0.14  &     0.16  &     0.13  &     0.04 \\
\species{Na}{i}      &     0.07  &  $-$0.02  &     0.01  &     0.08 \\
\species{Mg}{i}      &     0.02  &  $-$0.03  &     0.03  &     0.05 \\
\species{Al}{i}      &     0.06  &  $-$0.01  &  $-$0.02  &     0.06 \\
\species{Si}{i}      &  $-$0.06  &     0.06  &     0.07  &     0.04 \\
\species{Ca}{i}      &     0.10  &  $-$0.06  &     0.01  &     0.19 \\
\species{Sc}{ii}     &  $-$0.03  &     0.11  &     0.10  &     0.13 \\
\species{Ti}{i}      &     0.14  &     0.00  &  $-$0.01  &     0.10 \\
\species{Ti}{ii}     &  $-$0.04  &     0.11  &     0.10  &     0.11 \\
\species{V}{i}       &     0.17  &     0.10  &     0.00  &     0.20 \\
\species{Cr}{i}      &     0.11  &  $-$0.01  &     0.01  &     0.09 \\
\species{Cr}{ii}     &  $-$0.09  &     0.11  &     0.09  &     0.06 \\
\species{Mn}{i}      &     0.08  &  $-$0.08  &     0.09  &     0.11 \\
\species{Ni}{i}      &     0.12  &     0.06  &     0.08  &     0.13 \\
\species{La}{ii}     &     0.01  &     0.11  &     0.10  &     0.15 \\
\species{Eu}{ii}     &  $-$0.01  &     0.12  &     0.10  &     0.15 \\
\species{Li}{i}      &     0.15  &     0.00  &     0.00  &     0.00 \\
         C$_2$       &  $-$0.02  &     0.05  &     0.05  &     0.00 \\
            CN       &  $-$0.03  &     0.15  &     0.15  &     0.05 \\
$[$\species{O}{i}$]$ &     0.02  &     0.13  &     0.12  &     0.01 \\
\enddata
\end{deluxetable}
\end{center}

\end{document}